\documentclass[12pt,twoside,pointlessnumbers,nofootinbib,smallheadings]{revtex4}
\usepackage{amsmath}
\usepackage{amssymb}
\usepackage{color}

\usepackage{hangcaption,fancybox,feynmp}
\usepackage{indentfirst}
\usepackage{graphicx}
\usepackage{epsfig,subfigure,psfrag}
\usepackage{CJK}
\usepackage{mathrsfs}

\begin{document}


\title{
New color-octet axial vector boson revisited}
\author{Hao Wang $^{1}$, 
You-kai Wang $^{2,3}$ \footnote{wangyk@pku.edu.cn},
Bo Xiao $^{2}$, 
and Shou-hua Zhu $^{2,4}$ }

\affiliation{ $ ^1$ The Department of Astronomy, Beijing Normal
University, Beijing 100871, China \\
$ ^2$ Institute of Theoretical Physics $\&$ State Key Laboratory of
Nuclear Physics and Technology, Peking University, Beijing 100871,
China \\
$ ^3$ Institute of Theoretical Physics,
Chinese Academy of Sciences, Beijing 100190, China\\
$ ^4$ Center for High Energy Physics, Peking University,
Beijing 100871, China }

\date{\today}

\maketitle

\begin{center}
{\bf Abstract}

\begin{minipage}{15cm}
{\small  \hskip 0.25cm

In this paper we reexamine how to utilize the previous proposed
color-octet axial-vector boson $Z_{\rm C}$ to explain
the $3.4\sigma$ anomaly of $t\bar t$ forward-backward (FB) asymmetry $A_{\rm FB}$  for
$m_{t\bar t}> 450$GeV observed by CDF. Our numerical results indicate
 that the best-fit parameters are $g_A^q=0.07$, $g_A^Q=3$, and $M_{\rm C}=440\text{GeV}$, which
are obtained by fitting the mass dependent $A_{\rm FB}$ and total
cross section data provided by a recent CDF measurement. Here $g_A^q
(g_A^Q)$ and $M_{\rm C}$ are the axial couplings among $Z_{\rm C}$
with the first two (the third) generation quarks, and  $Z_{\rm C}$
mass, respectively. We also calculate one-side forward-backward
asymmetry $A_{\rm OFB}$ for top and bottom quark pair production at
the LHC, focusing on the new contributions from $Z_{\rm C}$. Our
studies show that $A_{\rm OFB}$ can be utilized  to measure the
properties of new particle $Z_{\rm C}$.
 }

\end{minipage}
\end{center}


\newpage
\section{Introduction\label{Introduction}}
 Recently, CDF and D0 measured the
$t\bar t$ forward-backward asymmetry at the Tevatron and found an
approximate $2\sigma$ deviation from the SM prediction
\cite{Abazov:2007qb,Aaltonen:2008hc,CDFnote9724,D0Note6062.CONF,
CDFnote10224,Kuhn:1998kw,Kuhn:1998jr,Antunano:2007da,Bernreuther:2010ny,
Bevilacqua:2010qb,Almeida:2008ug,Ahrens:2010zv,Ahrens:2011mw}.
 In the
latest CDF analysis \cite{Aaltonen:2011kc}, it is found that the
measured $A_{\rm FB}$ is larger than the SM prediction by
$3.4\sigma$ in the $M_{t\bar t}>450 \text{GeV}$ region. These
experiment results induced lots of new physics (NP) discussions
\cite{Jung:2009jz,Cao:2010zb,Bhattacherjee:2011nr,Jung:2011zv,
Shu:2011au,AguilarSaavedra:2011zy,Degrande:2011rt,Cao:2011ew,
Berger:2011ua,Cheung:2009ch,Shu:2009xf,Arhrib:2009hu,Dorsner:2009mq,
Cheung:2011qa,Gresham:2011dg,Barger:2011ih,Grinstein:2011yv,
Patel:2011eh,Ligeti:2011vt,Gresham:2011pa,Barger:2010mw,
Xiao:2010hm,Blum:2011up,Isidori:2011dp,Barreto:2011au,
Buckley:2011vc,Rajaraman:2011rw,Chen:2011mg,Nelson:2011us,
Jung:2011ua,Zhu:2011ww,Babu:2011yw,Cui:2011xy,Duraisamy:2011pt,Frampton:2009rk,
Ferrario:2010hm,Chivukula:2010fk,Martynov:2010ux,Choudhury:2010cd,
Bai:2011ed,Haisch:2011up,Djouadi:2009nb,Delaunay:2011vv,
Djouadi:2011aj,Barcelo:2011fw,Ferrario:2009bz,Ferrario:2009ee,
Rodrigo:2010gm,Bauer:2010iq,Xiao:2010ph,Wang:2011ta,Westhoff:2011ir,Zerwekh:2011wf,Shao:2011wa}.

Of all these NP explanations, one of the most discussed model is the
t-channel $Z^\prime$/$W^\prime$/scalar model
\cite{Jung:2009jz,Cao:2010zb,Bhattacherjee:2011nr,Jung:2011zv,Shu:2011au,AguilarSaavedra:2011zy,
Degrande:2011rt,Cao:2011ew,Berger:2011ua,Cheung:2009ch,Shu:2009xf,Arhrib:2009hu,Dorsner:2009mq,
Cheung:2011qa,Gresham:2011dg,Barger:2011ih,Grinstein:2011yv,Patel:2011eh,Ligeti:2011vt,
Gresham:2011pa,Barger:2010mw,Xiao:2010hm,Blum:2011up,Isidori:2011dp,Barreto:2011au,Buckley:2011vc,
Rajaraman:2011rw,Chen:2011mg,Nelson:2011us,Jung:2011ua,Zhu:2011ww,Babu:2011yw,Cui:2011xy,Duraisamy:2011pt},
where the NP particle $Z^\prime$/$W^\prime$/scalar induces a new
t-channel $q\bar q\rightarrow t\bar t$. New asymmetric cross sections
 come from both the interference of the NP t-channel diagram with
the SM gluon propagated s-channel diagram  and the self-conjugation
of the NP t-channel diagram. Contributions to the cross section from
the above two resources have opposite sign, so an approximate
cancelation of the new total cross sections can be achieved, which
is required by the $t\bar t$ total cross section measurement. The
t-channel neutral $Z^\prime$
model \cite{Jung:2009jz,Cao:2010zb,Bhattacherjee:2011nr,Jung:2011zv,Shu:2011au,AguilarSaavedra:2011zy,Degrande:2011rt}
is strongly restricted by the same-sign top quark pair production at
the Tevatron\cite{Aaltonen:2008hx,CDFnote10466} and the
LHC \cite{Collaboration:2011dk,Jung:2009jz,Cao:2011ew,Berger:2011ua,Jung:2011zv,Shu:2011au,Degrande:2011rt};
it is disfavored by the latest CMS \cite{Collaboration:2011dk}
measurement of the same-sign top quark pair production at the LHC.
$W^\prime$/scalar is not sensitive to the same-sign top quark pair
production, but can be easily tested (with only
$\mathcal{O}(\text{fb}^{-1})$ of integrated luminosity) at the LHC
though the $t\bar t+\text{jets}$ channel
\cite{Cheung:2009ch,Shu:2009xf,Arhrib:2009hu,Dorsner:2009mq,
Cheung:2011qa,Gresham:2011dg,Barger:2011ih,Grinstein:2011yv,
Patel:2011eh,Ligeti:2011vt,Gresham:2011pa,Jung:2011zv,Shu:2011au}.

Another kind of widely discussed NP is the s-channel color-octet
axial-vector
boson\cite{Frampton:2009rk,Ferrario:2010hm,Chivukula:2010fk,
Martynov:2010ux,Choudhury:2010cd,Bai:2011ed,Shu:2011au,
Haisch:2011up,Djouadi:2009nb,Delaunay:2011vv,Djouadi:2011aj,
Barcelo:2011fw,Ferrario:2009bz,Ferrario:2009ee,Cao:2010zb,
Rodrigo:2010gm,Bauer:2010iq,Xiao:2010ph,Wang:2011ta,Westhoff:2011ir},
where the new boson induces a new s-channel $q\bar q\rightarrow
t\bar t$. The new asymmetric contributions can possibly come from
the interference between the new s-channel diagram and the SM gluon
propagated s-channel diagram, and from the self-conjugation of the
new s-channel diagram. To generate positive extra asymmetric cross section, axial couplings with light and heavy quarks must have opposite sign ($g^q_A g_A^Q <0$). The axigluon model, where the couplings of axigluon to light quarks and heavy quarks are both at the strong
coupling
level \cite{Shu:2011au,Haisch:2011up,Djouadi:2009nb,Ferrario:2009bz,
Frampton:2009rk,Ferrario:2010hm,Rodrigo:2010gm,Martynov:2010ux,Choudhury:2010cd},
is strongly constrained by the dijet
\cite{Aaltonen:2008dn,Collaboration:2010eza,Khachatryan:2010jd,
Khachatryan:2011as,Aad:2011aj,Gresham:2011pa,Bai:2011ed,Shu:2011au,
Djouadi:2011aj,Barcelo:2011fw,Haisch:2011up,Choudhury:2010cd}
measurements; indeed, the axigluon model proposed in
\cite{Frampton:2009rk} is already excluded by the latest dijet
search by ATLAS \cite{Aad:2011aj,Shu:2011au}.

In the paper \cite{Xiao:2010ph}, part of us for the first
time proposed a $\sim 350\mbox{GeV}$ color-octet axial-vector boson
$Z_{\rm C}$ to explain the top quark $A_{\rm FB}$ anomaly. $Z_{\rm
C}$'s parameters are determined by the so called ``above" and
``below" mass dependent $A_{\rm FB}$ data, available at that time.
$Z_{\rm C}$ is different from the usually discussed
$\mathcal{O}(1\text{TeV})$ heavy axigluon
\cite{Frampton:2009rk,Ferrario:2010hm,Chivukula:2010fk,Martynov:2010ux,
Choudhury:2010cd,Bai:2011ed,Shu:2011au,Haisch:2011up} or KK-gluon
\cite{Djouadi:2009nb,Delaunay:2011vv,Djouadi:2011aj,Barcelo:2011fw}
or other color-octet resonance
\cite{Ferrario:2009bz,Ferrario:2009ee,Cao:2010zb,Rodrigo:2010gm,Bai:2011ed}
as its mass is near the top pair threshold. There are also some literatures \cite{Tavares:2011zg, Barcelo:2011fw,
Barcelo:2011vk, Alvarez:2011hi} which
propose about hundreds GeV new color-octet axial-vector particles to explain
the $A_{\rm FB}$ anomaly at the Tevatron. If the new particle's mass is still much
larger than the top pair mass threshold, such as $700\sim 800$ GeV in
Ref.\cite{Barcelo:2011fw, Barcelo:2011vk, Alvarez:2011hi} or
$\mathcal{O}(1\text{TeV})$ heavy axigluon, it must satisfy $g^q_A g^Q_A<0$ to generate
extra asymmetric cross sections. However, for the color-octet axial-vector
boson $Z_{\rm C}$ in our previous paper\cite{Xiao:2010ph} and $400\sim 450$GeV
axigluon in Ref.\cite{Tavares:2011zg}, the axial couplings to both light
and heavy quarks can have the same sign because of the light mass near
the top pair threshold. The difference between light axigluon in
Ref.\cite{Tavares:2011zg} and $Z_{\rm C}$ proposed by us is axigluon
in Ref.\cite{Tavares:2011zg} has flavor universal couplings to quarks
and $Z_{\rm C}$ has flavor non-universal couplings. In the paper
\cite{Wang:2011ta}, we studied the possible explanation of both top
$A_{\rm FB}$ and the dijet bump in the WZ/WW channel by adopting our
proposed color-octet axial-vector boson $Z_{\rm C}$ and put the
$Z_{\rm C}$'s mass $M_{\rm C}\sim 140\mbox{GeV}$. However we found
that such parameters can not perfectly account for
the recent CDF top $A_{\rm FB}$ mass dependent measurements,
especially for the $3.4\sigma$ deviation for $M_{t\bar t}> 450$GeV.
In this paper, we will focus on this issue and $Z_{\rm C}$'s
parameters are fitted and contour diagrams are drawn, possible cross
checks at the LHC are also discussed.

Our paper is organized as follows. Section \ref{sec.model} describes
the feature of the  $Z_{\rm C}$ model. In Section \ref{sec.analysis}
we check the feasibility of this model in explaining the experiment,
and obtain the constraints on the model parameters. Implications of
this $Z_{\rm C}$ model at the LHC is discussed in Sec. \ref{sec.lhc},
and we conclude in Sec. \ref{sec.conclusion}.

\section{Model Description\label{sec.model}}
The squared matrix element of the $q\bar q \rightarrow t\bar t$
process with mediating a SM gluon or $Z_{\rm C}$ is
\begin{equation}
 \begin{array}{rl}
 \sum\limits_{\text{Color, Spin}}|M|^2=& \frac{C_A C_F}{2}\{4g_s^4 (1+c^2+4m^2)+\frac{8g_s^2
 \hat{s}(\hat{s}-M_{\rm C}^2)}{(\hat{s}-M_{\rm C}^2)^2+M_{\rm C}^2\Gamma_{\rm C}^2} [g_V^q g_V^t(1+c^2+4m^2)+2g_A^q g_A^t c]\\\\
 &+\frac{4\hat{s}^2}{(\hat{s}-M_{\rm C}^2)^2+M_{\rm C}^2\Gamma_{\rm C}^2}\left[8 g_V^q g_A^q g_V^t
 g_A^t
 c+((g_V^q)^2+(g_A^q)^2)\times\right.\\\\
 &\left.((g_V^t)^2(1+c^2+4m^2)+(g_A^t)^2(1+c^2-4m^2))\right]\},
 \end{array}
\label{axigluon}
\end{equation}
where $q$ represents any of the light quarks;
$m=m_t/\sqrt{\hat{s}}$, $\beta=\sqrt{1-4m^2}$, $c=\beta \cos\theta$;
and  $g_V^{q (t)}/g_A^{q (t)}$ are vector- and axial-vector
couplings among light quarks (top) and $Z_{\rm C}$. $\Gamma_{\rm C}$ is
the width of $Z_{\rm C}$. Terms on the right-hand side represent QCD
amplitude self-conjugation, the interference between QCD and $Z_{\rm
C}$ amplitudes and $Z_{\rm C}$ amplitude self-conjugation,
respectively.

Equation (\ref{axigluon}) indicates that only odd $c$ terms can
contribute to the forward-backward asymmetry. To suppress the
impact on the total cross section, it is reasonable to require
the vectorlike couplings $g_V^q=g_V^t=0$. So the new boson has a
pure axial-vector coupling to the quarks
\cite{Ferrario:2009ee,Cao:2010zb,Ferrario:2010hm,Choudhury:2010cd,Bai:2011ed,Barcelo:2011fw,Xiao:2010ph,Wang:2011ta}.
Under this assumption, Eq. (\ref{axigluon}) now becomes,
\begin{equation}
 \begin{array}{rl}
 \sum\limits_{\text{Color, Spin}}|M|^2=& \frac{C_A C_F}{2}\{4g_s^4 (1+c^2+4m^2)+\frac{8g_s^2
 \hat{s}(\hat{s}-M_{\rm C}^2)}{(\hat{s}-M_{\rm C}^2)^2+M_{\rm C}^2\Gamma_{\rm C}^2}2g_A^q g_A^t c\\\\
 & +\frac{4\hat{s}^2}{(\hat{s}-M_{\rm C}^2)^2+M_{\rm C}^2\Gamma_{\rm C}^2}
 (g_A^qg_A^t)^2(1+c^2-4m^2)\},
 \end{array}
\label{eq.simp_axigluon}
\end{equation}
in which the first term is the SM gluon mediated contribution, the
second term is the interference between SM and $Z_{\rm C}$ process,
which contributes to a nonezero asymmetric cross section, and the
third term is the self-conjugation of the $Z_{\rm C}$ process, which
may contribute to the total cross section.

From Eq. (\ref{eq.simp_axigluon})  one can see clearly that the
product $g_A^q g_A^t$ must be large enough to generate an extra
asymmetric cross section. On the other hand, $g_A^q$ must be small
in order to eliminate the extra contribution to the heavy quark and
light dijet cross sections. So the axial couplings can be assumed
as,
\begin{equation}
 \begin{split}
 g_A^u=g_A^d=g_A^c=g_A^s\equiv g_A^q&\\
 g_A^b=g_A^t\equiv g_A^Q&
 \end{split}\;\;\;\;
 \text{with}\;\;\;\;g_A^q< g_A^Q.
 \label{eq.model_assumption}
\end{equation}
Therefore \{$g_A^q, g_A^Q, M_{\rm C}$\} then form the complete free
parameters set of the $Z_{\rm C}$ model. Note that here the third
generation quarks now have an universal coupling $g_A^Q$ with
$Z_{\rm C}$, which is different from the assumption in our previous papers
\cite{Xiao:2010ph,Wang:2011ta}. In \cite{Xiao:2010ph,Wang:2011ta},
$Z_{\rm C}$'s coupling with bottom quark is set to be the same as
those with the first two generation quarks. It will be convenient
for the model construction for a generic third-generation coupling,
as well as a larger coupling with bottom quarks will broaden $Z_{\rm
C}$'s width $\Gamma_{\rm C}$, which makes it easier to be hidden in the
invariant mass spectrum. $g^b_A$ may be constrained from $b$
physics, however, it can be expected that such constraints would be
moderate due to the huge QCD backgrounds. There will be a brief
estimation based on the optimal fitted parameters in the following sections.

We shall check the feasibility of this model in explaining the
latest experiments, and explore the experimental constraints on
these parameters. This will be the content of the next section.

\section{Analysis}\label{sec.analysis}
As mentioned in the above section, there are three independent
parameters in the $Z_{\rm C}$ model, \{$g_A^q, g_A^Q, M_{\rm C}$\}.
These parameters can be constrained by comparing experimental
variables and their theoretical expected values at the Tevatron.
Here the variables are adopted as the top pair total cross section
$\sigma^{\rm tot}$, $(A_{{\rm FB}})_{b}$ with $M_{t\bar{t}}$ below
450GeV and $(A_{{\rm FB}})_{a}$ with $M_{t\bar{t}}$ above 450GeV.

Their explicit expressions can be listed as follows:
\begin{itemize}
\item{
Total cross section is a sum of the SM QCD and $Z_{\rm C}$ induced
cross sections.
\begin{equation}
\sigma^{\rm SM+Z_{\rm C}}=\sigma^{\rm SM}+\sigma^{Z_{\rm C}},
\end{equation}
Here $\sigma^{Z_{\rm C}}$ is taken as its born level expression, as
shown in the third term in Eq. (\ref{eq.simp_axigluon}).
$\sigma^{\rm SM}$ is up to NLO QCD level.  }
\item{
The asymmetry is contributed from both $Z_{\rm C}$ and SM QCD
resources.
\begin{equation}
\begin{array}{rl}
(A_{\rm FB}^{\rm SM+Z_{\rm
C}})_{a/b}&=\frac{(\sigma_A^\text{SM})_{a/b}+(\sigma_A^{Z_{\rm
C}})_{a/b}}{(\sigma^{\text{SM}})_{a/b}+(\sigma^{Z_{\rm C}})_{a/b}}
=\frac{(\sigma_A^\text{SM})_{a/b}+(\sigma_A^{Z_{\rm
C}})_{a/b}}{(\sigma^\text{SM})_{a/b}}
\frac{(\sigma^\text{SM})_{a/b}}{(\sigma^\text{SM})_{a/b}+(\sigma^{Z_{\rm
C}})_{a/b}}\\ \\
& \approx\left[(A_{\rm FB}^{\rm SM})_{a/b}+\frac{(\sigma_A^{Z_{\rm
C}})_{a/b}}{(\sigma^\text{LO})_{a/b}}\right]
\frac{(\sigma^\text{LO})_{a/b}}{(\sigma^\text{LO})_{a/b}+(\sigma^{Z_{\rm
C}})_{a/b}},
\end{array}
\end{equation}
where $\sigma_A$ is the asymmetric cross section, and the subscript
$a$/$b$ denotes that the $M_{t\bar t}$ are integrated above/below
than 450GeV. Note that in the experiments, the SM expectation of the
asymmetry is obtained from Monte Carlo generators, in which the
denominator is taken as the NLO QCD cross section. It will be
different with the usual theoretical calculation with a $K$
factor \cite{Wang:2010du}, which is at order $K\sim 1.3$. We neglect
this effect here. }
\end{itemize}

For all the three variables, their one standard deviations are taken
as the corresponding experimental errors. Some relative quantities
are listed in Table \ref{compareSM-Ex}.

\begin{table}[htb]
\caption{\label{compareSM-Ex}Relative experimental results and SM
expectations\cite{Aaltonen:2010ic,Aaltonen:2011kc} to obtain
$\sigma^{\rm tot}$ and $(A_{\rm FB})_{a/b}$.}
\begin{tabular}{cccccc}
\hline\hline
 & $\sigma^{\rm tot}$ & $(A_{\rm FB})_{b}$ & $(A_{\rm FB})_{a}$ & $(\sigma^{\rm LO})_{b}$ & $(\sigma^{\rm LO})_{a}$\\
\hline
EXP & $7.70 \pm 0.52$pb & $ - 0.116 \pm {\rm{0}}{\rm{.153}}$ & $0.475 \pm {\rm{0}}{\rm{.114}}$ & $\cdots$ & $\cdots$\\
\hline
SM  & $7.45_{ - 0.63}^{ + 0.72}$pb & $0.040 \pm 0.006$ & $0.088 \pm 0.013$ & 3.70pb & 2.23pb\\
\hline\hline
\end{tabular}
\end{table}

In our numerical calculation, the SM parameters are set as
\begin{equation}
\label{eq.sm_para} \alpha_s(m_Z)=0.118, m_t=171.2\text{GeV},
m_b=4.7\text{GeV}.
\end{equation}
the renormalization and factorization scales are chosen as
$\mu_R=\mu_F=m_t$; the PDF package CTEQ6L is used for the LO
calculation and CTEQ6m is used for the NLO QCD calculation.

Generally speaking, there are two alternative methods to find the
possible parameter region for $M_{\rm C}$. The first approach is to
consider the constraint one by one independently, the second approach is
to construct a total $\chi^2$ by utilizing all inputs.
We will use both methods in the following studies. It will be seen later that the two
methods give the consistent results.

\begin{figure}[htbp]
\begin{center}
\includegraphics[width=0.80\textwidth]
{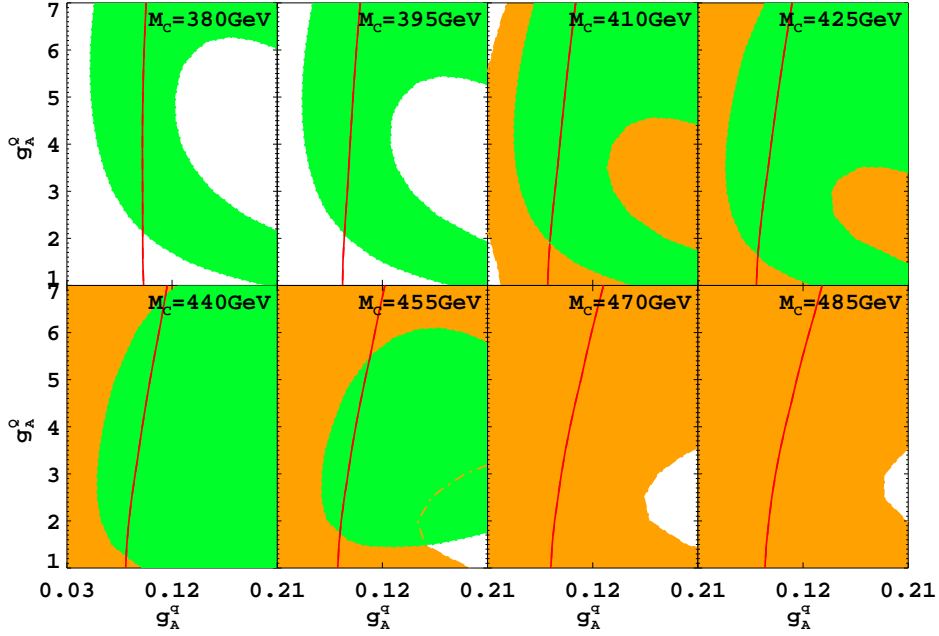}
\end{center}
\caption{\label{Deviation}The deviation contours on the
$g_A^q-g_A^Q$ plane with different $M_{\rm C}$s.}
\end{figure}

Figure \ref{Deviation} shows the contour diagram of the three
independent constraints with $M_{\rm C}$ varying from 380 GeV to
485 GeV with a step of 15 GeV. The green/yellow area is the 1 $\sigma$
allowed parameter region by requirement of the $A_{\rm FB}$
above/below $M_{t\bar{t}}=450~\mbox{GeV}$. The area in the left
region of the red curve is the allowed region from constraint from
$t\bar{t}$ total cross section. The constraints can be understood by
referring Eq. (\ref{eq.simp_axigluon}): For a lighter $Z_{\rm C}$
with its mass smaller than 450 GeV, it will be easy to induce the
measured $A_{\rm FB}$ for $M_{t\bar{t}}>450~\mbox{GeV}$ but difficult
for $M_{t\bar{t}}<450~\mbox{GeV}$. The reason is that $\hat{s}-M_{\rm
C}^2$ is more likely to be positive for
$M_{t\bar{t}}<450~\mbox{GeV}$, opposite to the measured central value
of $A_{\rm FB}$. The opposite situation will happen when $M_{\rm C}$ is
larger than $450~\mbox{GeV}$, where $\hat{s}-M_{\rm C}^2$ is likely
not large enough to produce enough asymmetric cross section and the
product $g^q_A g^Q_A$ should be lager, out of the plotted region in
Fig.~\ref{Deviation}. $Z_{\rm C}$'s impact on the total cross section
will be always small, as can be seen in the third term in Eq.
(\ref{eq.simp_axigluon}), so this constraint is not sensitive to the
variation of the three parameters. After applying the three
constraints, the overlapping region is $410~\text{GeV}\lesssim M_{\rm
C}\lesssim 455~\text{GeV}$.

\begin{figure}[htbp]
\begin{center}
\includegraphics[width=0.32\textwidth]
{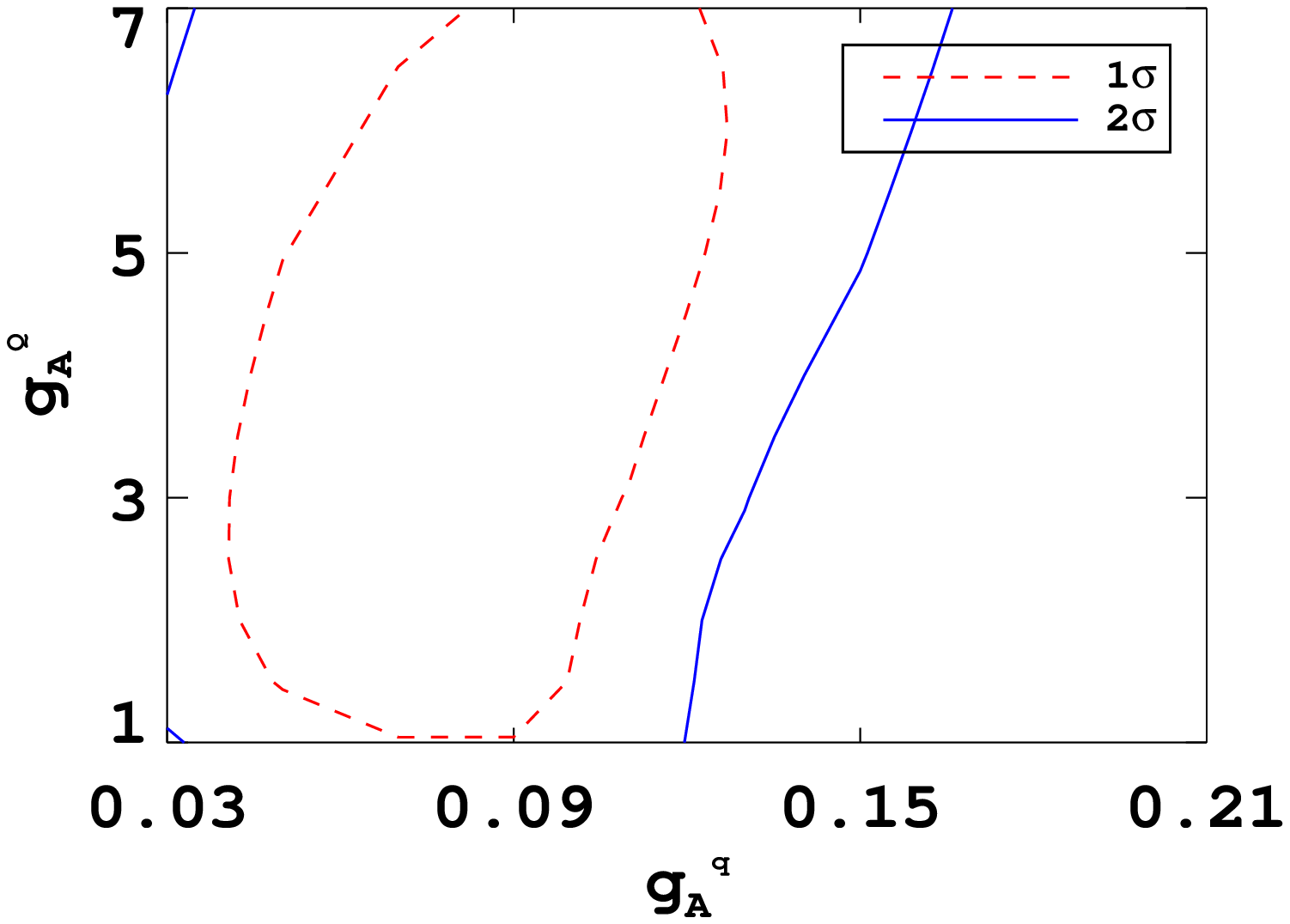}
\includegraphics[width=0.32\textwidth]
{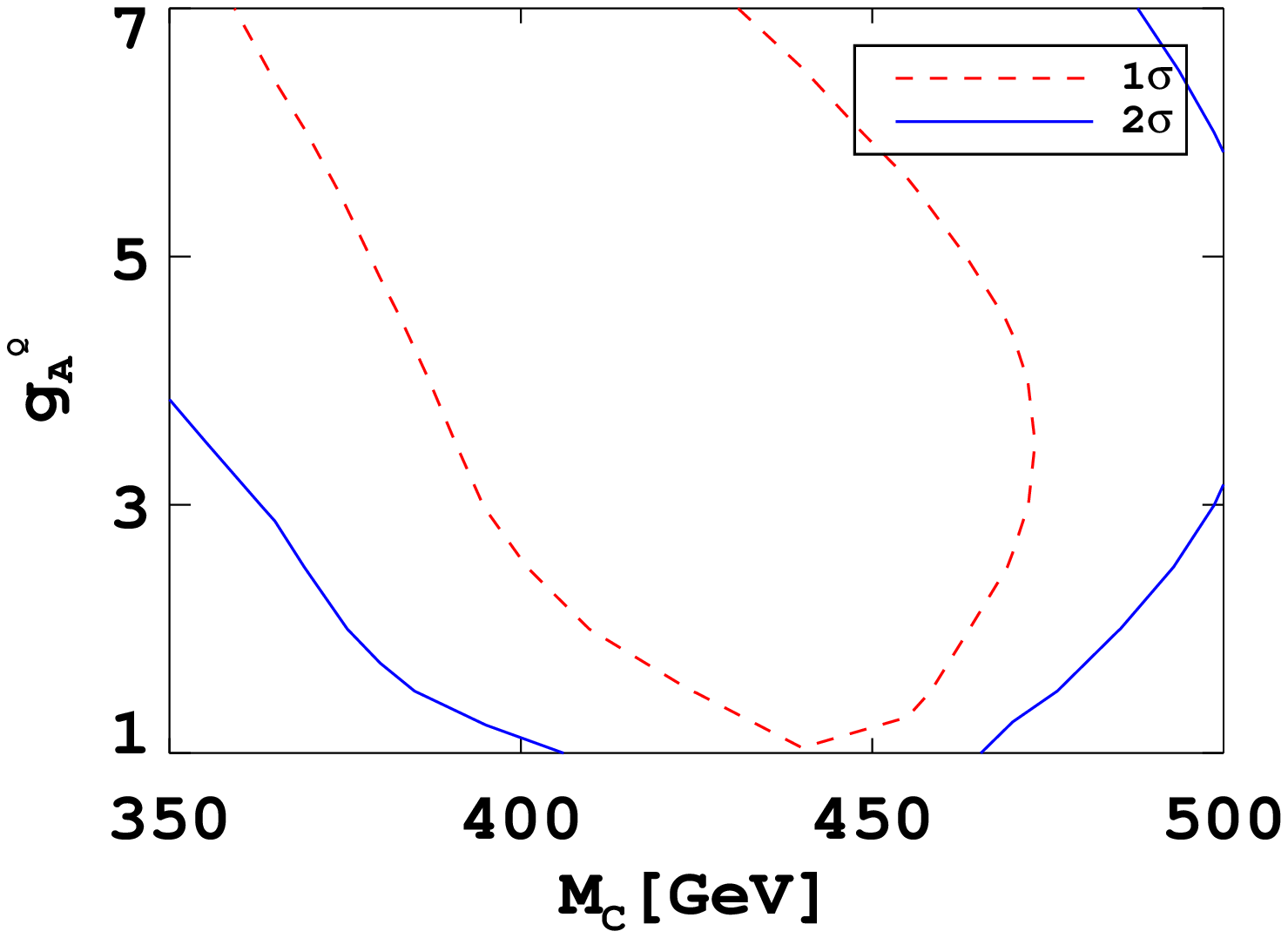}
\includegraphics[width=0.32\textwidth]
{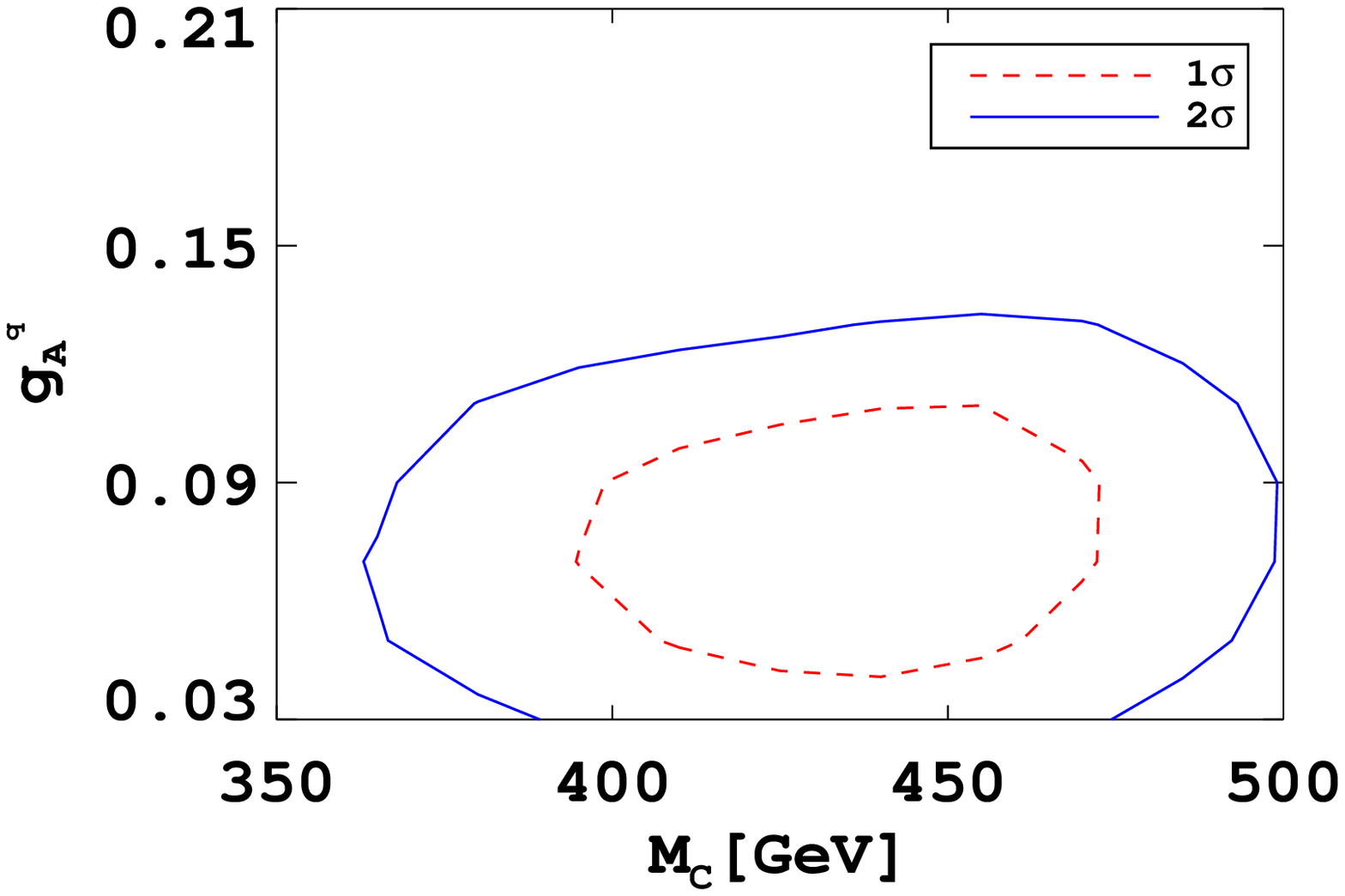}
\end{center}
\caption{\label{confidence}Two-dimensional 1$\sigma$(red) and
2$\sigma$(blue) confidence regions of the new $Z_{\rm C}$ model,
with the other one parameter fixed at its optimal point.}
\end{figure}

The second method is the standard $\chi^2$ fit, in which the
$\chi^2$ is defined as
\begin{equation}
 \chi^2\equiv \sum_{i}\frac{(O_i^{\rm th}-O_i^{\rm exp})^2}{(\delta O_i)^2},
 \label{eq.chi2}
\end{equation}
where $O_i$ represents the three observables $\sigma^{\rm tot}$,
$(A_{\rm FB})_{a}$ and $(A_{\rm FB})_{b}$. $\delta O_i$ are taken as
the corresponding experimental errors. A possible 3-dimension
parameter  region is scanned to find the minimal $\chi^2$ point.
Contour diagrams are obtained by the variation $\Delta
\chi^2\equiv\chi^2-\text{min}(\chi^2)$. In principle, such a
$\chi^2$ fit is not very suitable as there are too many free
parameters and too few data samples. However, we make the $\chi^2$
fit anyway and the situation may be improved in case of more
statistics in the future. Figure~\ref{confidence} shows the two-dimensional
contour diagrams with the other one parameter fixed at
its optimal point. The best-fit parameters are $M_{\rm
C}=440~\text{GeV}$, $g_A^Q=3.0$ and $g_A^q=0.07$.\footnote{By adopting
these optimal parameters, we estimated the impact on $R_b$ caused by
$Z_{\rm C}$ in $Z$ decay  according to formulas in Ref.\cite{Haisch:2011up}.
The vertex $Zb\bar{b}$ has about $0.4\%$ correction and the corrected $R_b$ agrees
with SM predicted value within 1.2 standard deviation. } Limited by the
accuracy of the Tevatron experimental data, large parameter space
regions can still survive. For $M_{\rm C}$, it can vary from
390~GeV to 470~GeV within 1 $\sigma$ deviation. $M_{\rm C}$ is
somewhat greater than the top pair threshold as the central value of
$(A_{\rm FB})_b$ is negative. $Z_{\rm C}$'s axial-vector like
coupling to the light quarks $g_A^q$ is about ($0.04\sim 0.12$).
The dijet constraints can be easily satisfied
\cite{Buckley:2011vc,Aaltonen:2008dn,Collaboration:2010eza,
Khachatryan:2010jd,Khachatryan:2011as,Aad:2011aj} as comparing to
the SM $Z$ boson's couplings to the light quark $\sim0.36$.
Figure~\ref{Deviation} shows that $g^Q_A$ must be large, which indicates
that $Z_{\rm C}$ maybe a condensate of the heavy quark pairs.
Here we take $g^Q_A$ as an effective coupling so the
born level calculation in the $Z_{\rm C}$ model is still reliable.

\begin{figure}[htbp]
\begin{center}
\includegraphics[width=0.7\textwidth]
{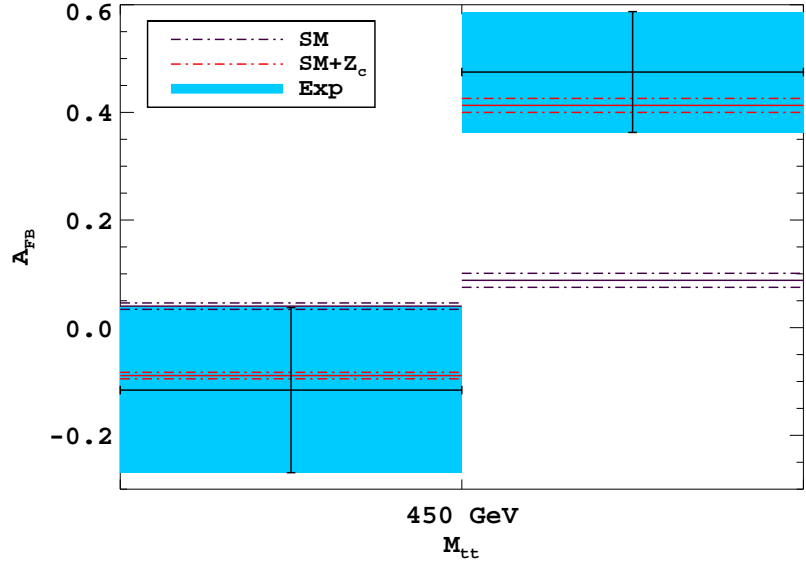}
\end{center}
\caption{\label{Afb-L-G}Comparison of the mass dependent $A_{\rm
FB}$ between the Tevatron experimental data\cite{Aaltonen:2009iz},
SM predictions and their theoretical value with the best-fitted
$Z_{\rm C}$ parameters. Exact numbers can be found in Table
\ref{compareSM-Ex} and Table \ref{compareZc-Ex}.}
\end{figure}

\begin{figure}[htbp]
\begin{center}
\includegraphics[width=0.7\textwidth]
{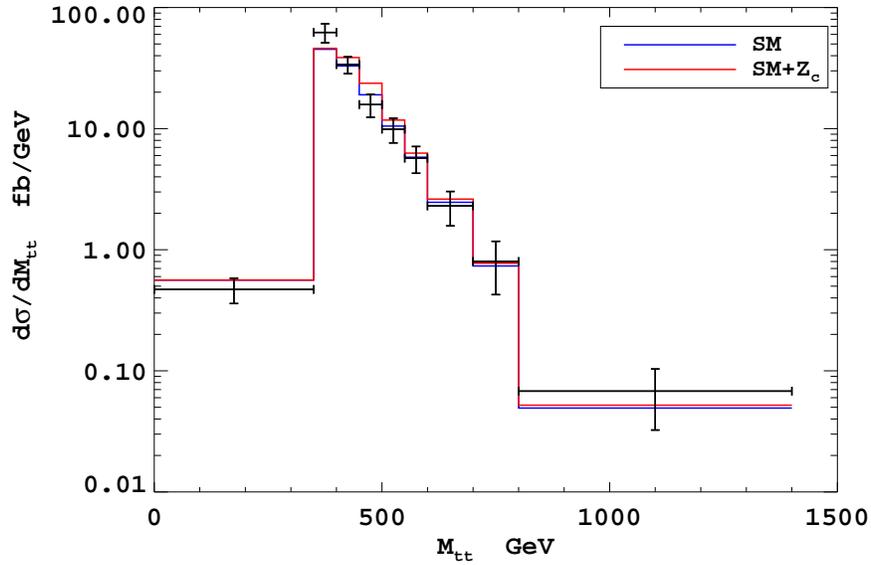}
\end{center}
\caption{\label{differential-CS} Small bump near 450~GeV induced by
$Z_{\rm C}$ on the $d\sigma/dM_{t\bar t}$ histograms. Tevatron
experimatal data are taken from \cite{Aaltonen:2009iz}.
The SM predicted values are up to NLO QCD.  }
\end{figure}

The expectations of the best-fit $Z_{\rm C}$ model, as well as the
corresponding experimental measurements, for the mass dependent
$A_{\rm FB}$ and the total cross section are listed in Table
\ref{compareZc-Ex}. Comparing with Table \ref{compareSM-Ex}, one can
see that by introducing $Z_{\rm C}$, the fit improves greatly and
the anomaly between the theory and the experiment disappear, which
are also illustrated in Fig.\ref{Afb-L-G}. Figure~\ref{differential-CS}
shows the impact of $Z_{\rm C}$ on the $d\sigma/dM_{t\bar t}$
distribution. A slight bump is introduced in the $d\sigma/dM_{t\bar
t}$ distribution. However, due to its small size and the
experimental uncertainty, this bump is hard to detect.

\begin{table}[htb]
\caption{\label{compareZc-Ex}The experimental data and the $Z_{\rm
C}$ model expectations of the total cross section, the $A_{\rm FB}$
in low and high invariant mass regions at the Tevatron.}
\begin{tabular}{cccc}
\hline\hline
 & $\sigma$[pb] & $(A_{\rm FB})_b$ & $(A_{\rm FB})_a$ \\
\hline
EXP & $7.70 \pm 0.52$ & $ - 0.116 \pm {\rm{0}}{\rm{.153}}$ & $0.475 \pm {\rm{0}}{\rm{.114}}$ \\
\hline
SM+$Z_{\rm C}$ & $8.12_{ - 0.63}^{ + 0.72}$ & $-0.089 \pm 0.006$ & $0.413 \pm 0.013$ \\
\hline\hline
\end{tabular}
\end{table}

\section{the implications of $Z_{\rm C}$ at the LHC}\label{sec.lhc}
The top quark $A_{\rm FB}$ anomaly discovered at the Tevatron can be
successfully explained by introducing the color-octet axial-vector-like
boson $Z_{\rm C}$. In this section, we will discuss how to study $Z_{\rm C}$
at the more powerful machine LHC.
Discussions will be focused on on two kinds of final states,
$t\bar t$ and $b\bar b$, as $Z_{\rm C}$ is assumed to couple
strongly to third generation quarks.

Different from the $p\bar{p}$ collider Tevatron, there is no
preferred  direction at the charge-symmetric $pp$ collider LHC.
Various definitions of forward-backward asymmetry are
proposed
\cite{Langacker:1984dc,Dittmar:1996my,Petriello:2008zr,Diener:2009ee,Diener:2010sy,Kuhn:1998kw,Kuhn:1998jr,Antunano:2007da,Ferrario:2008wm,Rodrigo:2008qe,Wang:2010du,Wang:2010tg,Xiao:2011kp}
to solve this problem. We make use of the so-called ``one-side
forward-backward asymmetry", $A_{\rm
OFB}$\cite{Wang:2010du,Wang:2010tg}, which is both conceptually
transparent and observationally easy to detect.

The definition of $A_{\rm OFB}$ is:
\begin{equation}
A_{\rm OFB}=\frac{F_- + B_-}{F_+ +B_+}\equiv\frac{\sigma^A}{\sigma},
\label{AFB2}
\end{equation}
with
\begin{equation} F_\pm= \left. \left(\sigma( \Delta Y>0)\pm \sigma(\Delta
Y<0)\right)\right|_{P_{Q\bar{Q}}^z>P_{\rm cut}^z}
\end{equation}
\begin{equation} B_\pm= \left. \left(\sigma(\Delta
Y<0)\pm \sigma(\Delta Y>0)\right)\right|_{P_{Q\bar{Q}}^z<-P_{\rm
cut}^z}
\end{equation}
where $\Delta Y=Y_Q-Y_{\bar Q}$ is the difference of rapidity
between $Q$ and ${\bar Q}$. $P_{\rm cut}^z$ is a cut on the
longitudinal momentum $P_{Q\bar{Q}}^z$. The detailed
definition can be found in Ref. \cite{Wang:2010du}.

Through our calculation, the SM parameters and PDF sets are chosen
as in the last section (see Eq.~(\ref{eq.sm_para}) and the context
therein), b-jet cut is taken as the transverse momentum
$P^b_T>20~\text{GeV}$ and $Y<2.5$, and the $t\bar t$ and $b\bar b$
detection efficiency are set as $\epsilon_{t\bar t}=4\%$ and
$\epsilon_{b\bar b}=25\%$, respectively \cite{Godfrey:2008vf}. The
energy of the LHC is set to be 7~TeV and an integrated luminosity of
10~fb$^{-1}$ is assumed.

\begin{figure}[htbp]
\begin{center}
\includegraphics[width=0.47\textwidth]
{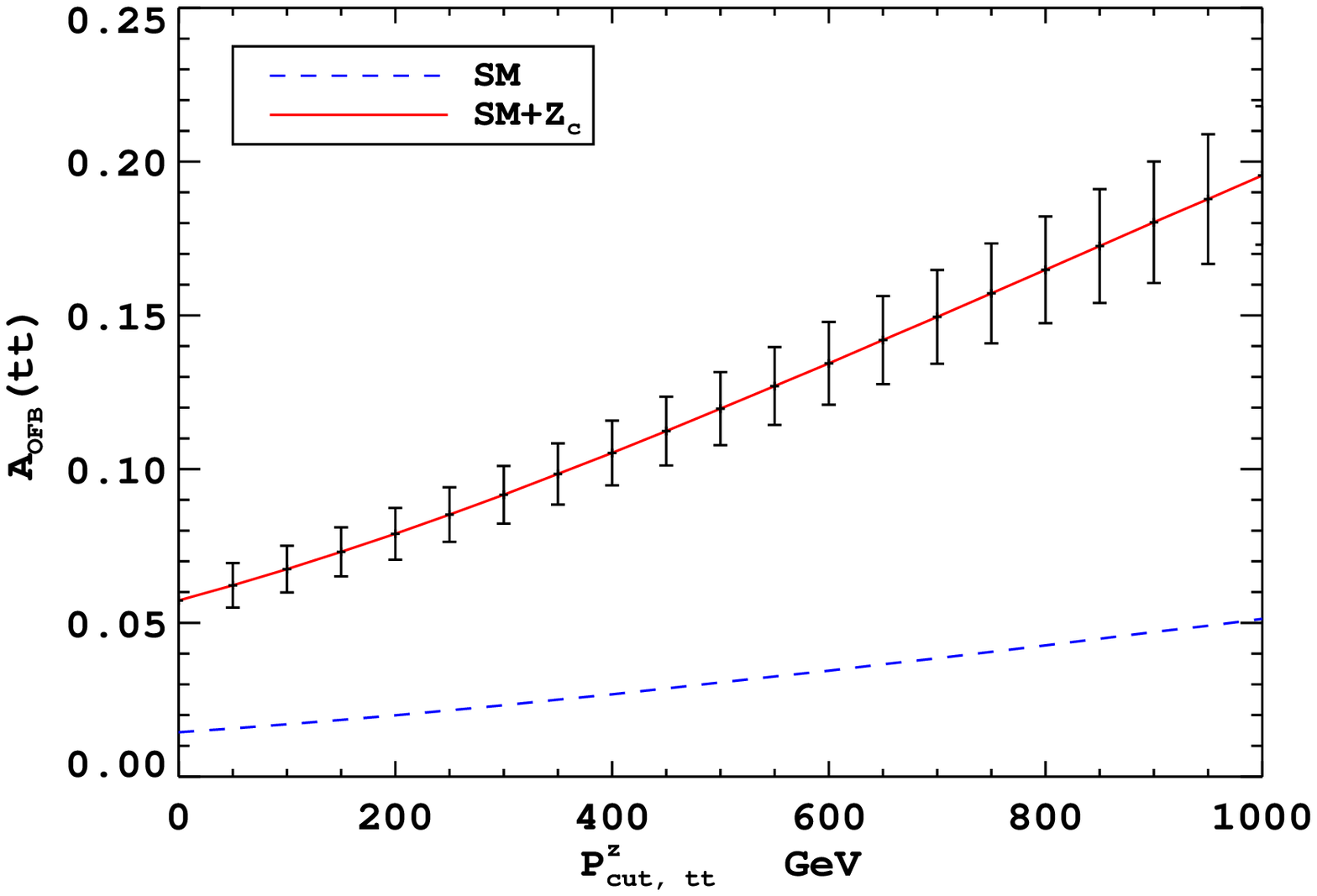}
\end{center}
\begin{center}
\includegraphics[width=0.47\textwidth]
{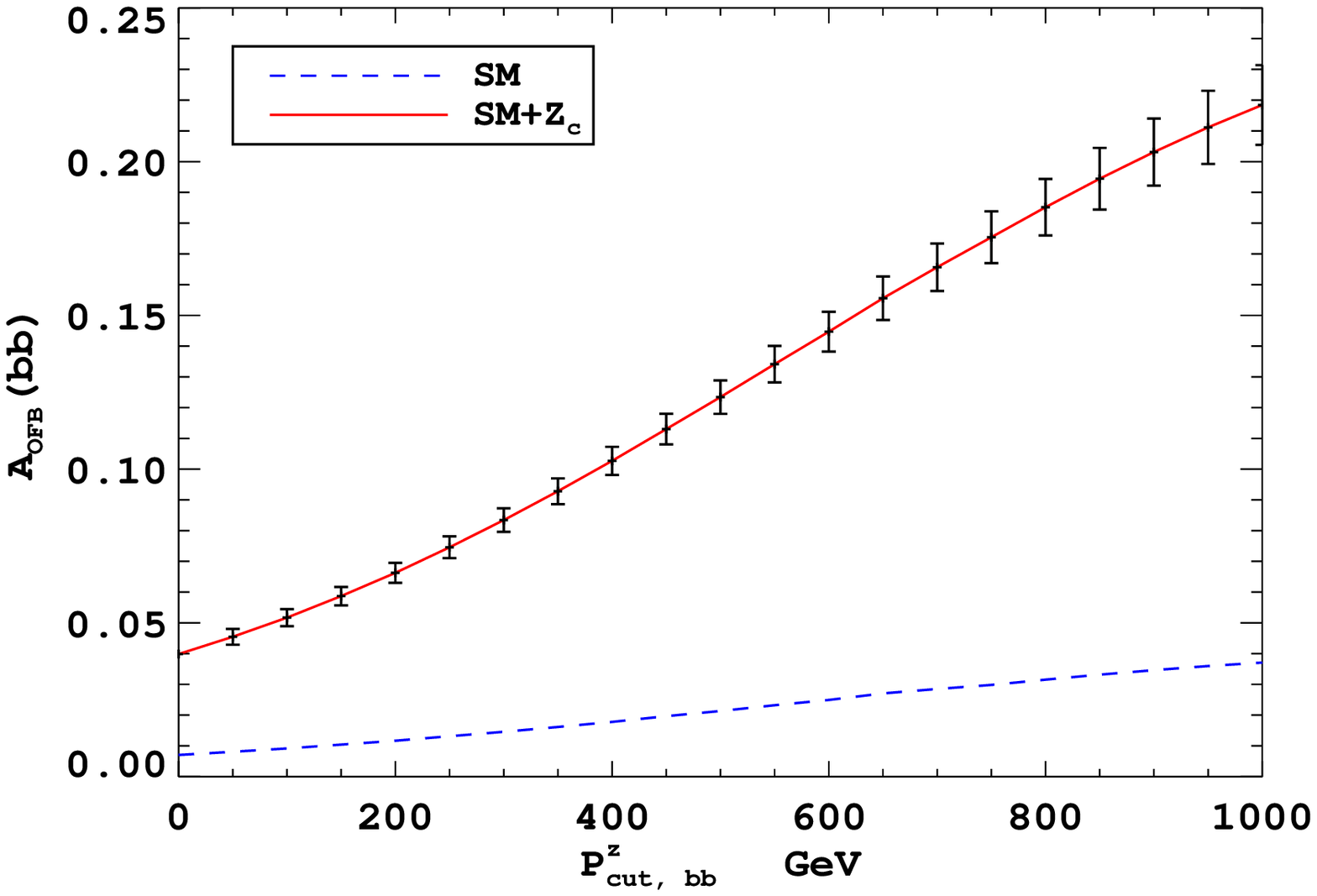}
\includegraphics[width=0.47\textwidth]
{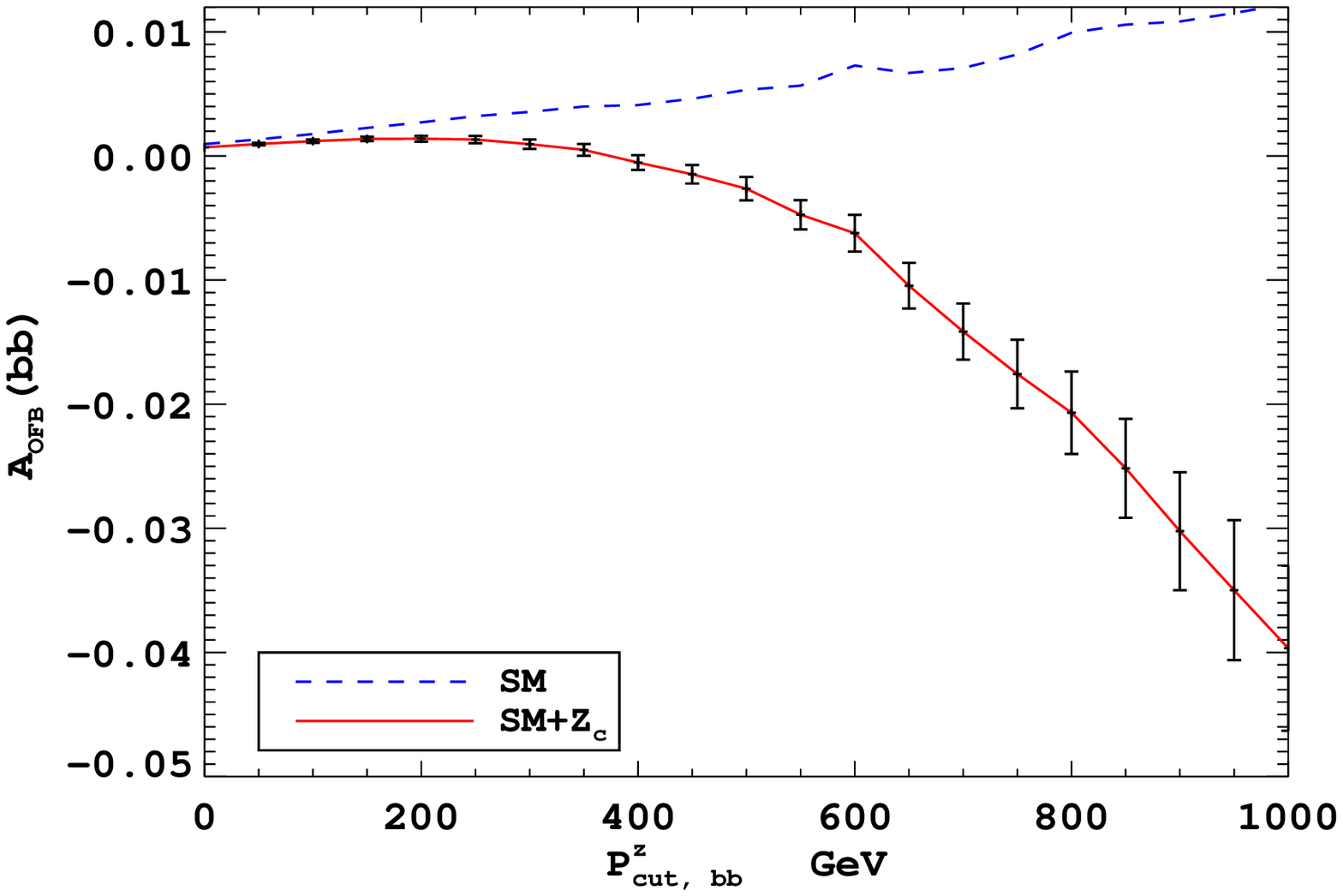}
\end{center}
\caption{\label{Afb-tt-bb-LHC}SM and SM+$Z_{\rm C}$ expectations of
the $A_{\rm OFB}(t\bar t)$ and $A_{\rm OFB}(b\bar b)$ as functions
of the $P^z_{{\rm cut},t\bar t}$ at the 7~TeV LHC. In the top panel
the $M_{t\bar t}>440~\text{GeV}$ cut is implemented; in the bottom
left (right) panel the $M_{b\bar b}>440~\text{GeV}$
($100~\text{GeV}<M_{b\bar b}<440~\text{GeV}$) cut is implemented.}
\end{figure}

$A_{\rm OFB}$ as a function of $P_{\rm cut}^z$ for $t\bar{t}$ and
$b{\bar b}$ final states are drawn in Fig.~\ref{Afb-tt-bb-LHC}.
$Z_{\rm C}$'s parameters are taken as their optimal values, $M_{\rm
C}=440~\text{GeV}$, $g_A^Q=3.0$ and $g_A^q=0.07$. In order to exhibit
the positive and negative values of $A_{\rm OFB}$, we show the
predicted $A_{\rm OFB}$ for $M_{Q\bar Q}> M_{\rm C}$ and $M_{Q\bar
Q}< M_{\rm C}$ respectively. According to the error propagation
formula, the statistical fluctuation of $A_{\rm OFB}$ can be
expressed as
\begin{equation}
 \delta A\equiv \sqrt{\frac{4N^FN^B}{N^3}}\simeq
 \frac{1}{\sqrt{\mathcal{L}\sigma\epsilon_{f\bar f}}},
\label{Afb-error}
\end{equation}
where $N^F/N^B$ are the number of forward/backward events, and
$N=N^F+N^B$ is the total events number. Error bars in
Fig.~\ref{Afb-tt-bb-LHC} stand only for statistical uncertainties. It
shows clearly that SM predictions are shifted significantly by
effects of $Z_{\rm C}$.

As mentioned in the above section, $Z_{\rm C}$ can cause a negative
sign $A_{\rm FB}$ in the $M_{Q\bar{Q}}<M_{\rm C}$ region, compared
to the always positive  $A_{\rm FB}$ for all energy regions in the
SM. This is an interesting signature for $t\bar t$ or $b\bar b$
final states. Figure~\ref{confidence} shows that $M_{\rm C}$ can
vary in an interval about $360\sim500$~GeV within 2 standard
deviations. If $M_{\rm C}>2m_t$, both $A_{\rm FB}(t\bar t)$ and
$A_{\rm FB}(b\bar b)$ will be negative in the interval
$2m_t<M_{Q\bar{Q}}<M_{\rm C}$. If $2m_b<M_{\rm C}<2m_t$, $A_{\rm
FB}(t\bar t)$ will be always positive and $A_{\rm FB}(b\bar b)$ will
be negative in the interval $2m_b<M_{Q\bar{Q}}<M_{\rm C}$. This
behavior can be used in checking the universal couplings of $Z_{\rm
C}$ with the third-generation quarks. So it is crucial to measure
the invariant mass dependent $A_{\rm FB}$ for both top and bottom
quark states to fix the location of the mass of $Z_{\rm C}$.

\begin{figure}[htbp]
\begin{center}
\includegraphics[width=0.47\textwidth]
{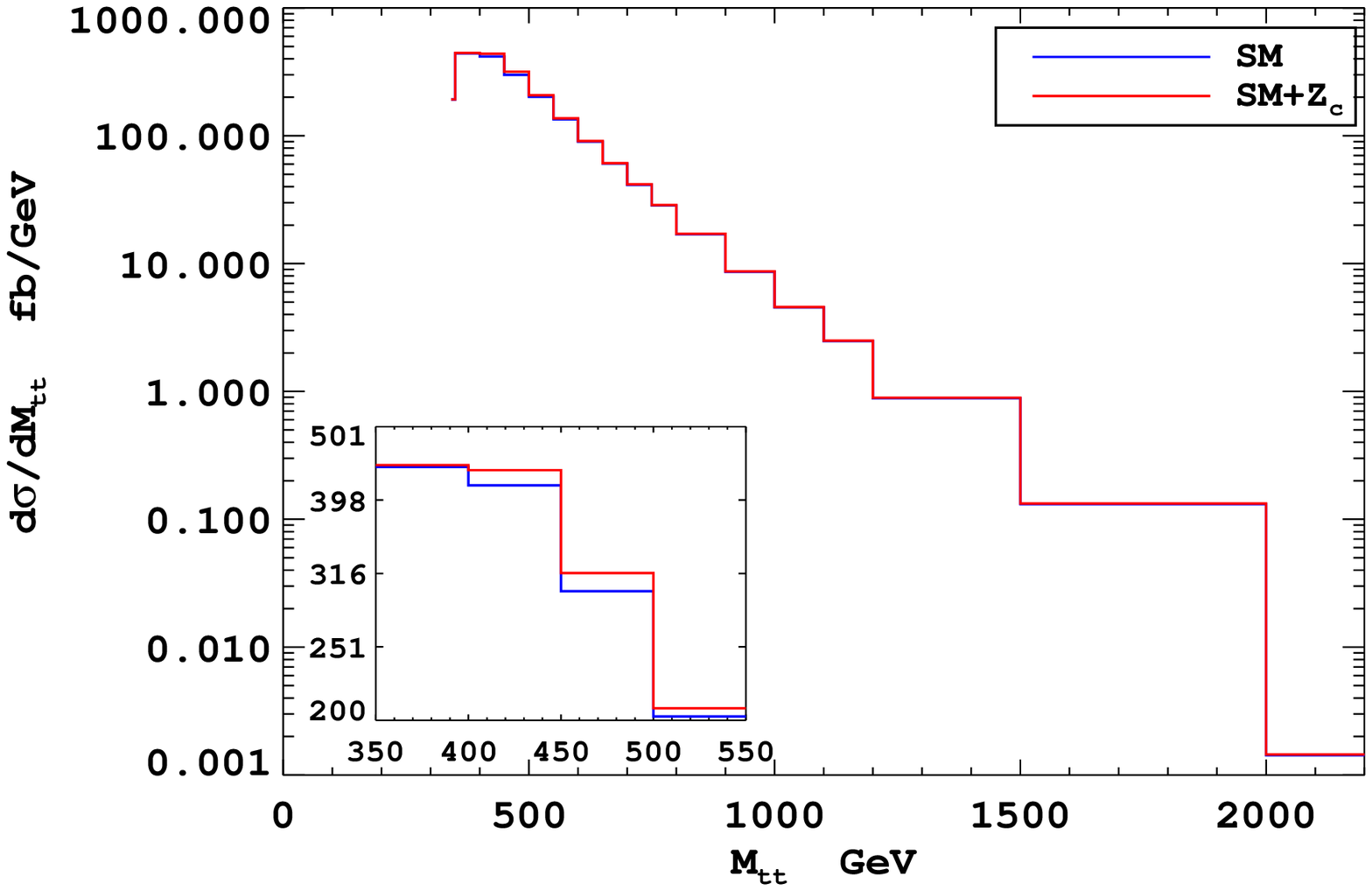}
\includegraphics[width=0.47\textwidth]
{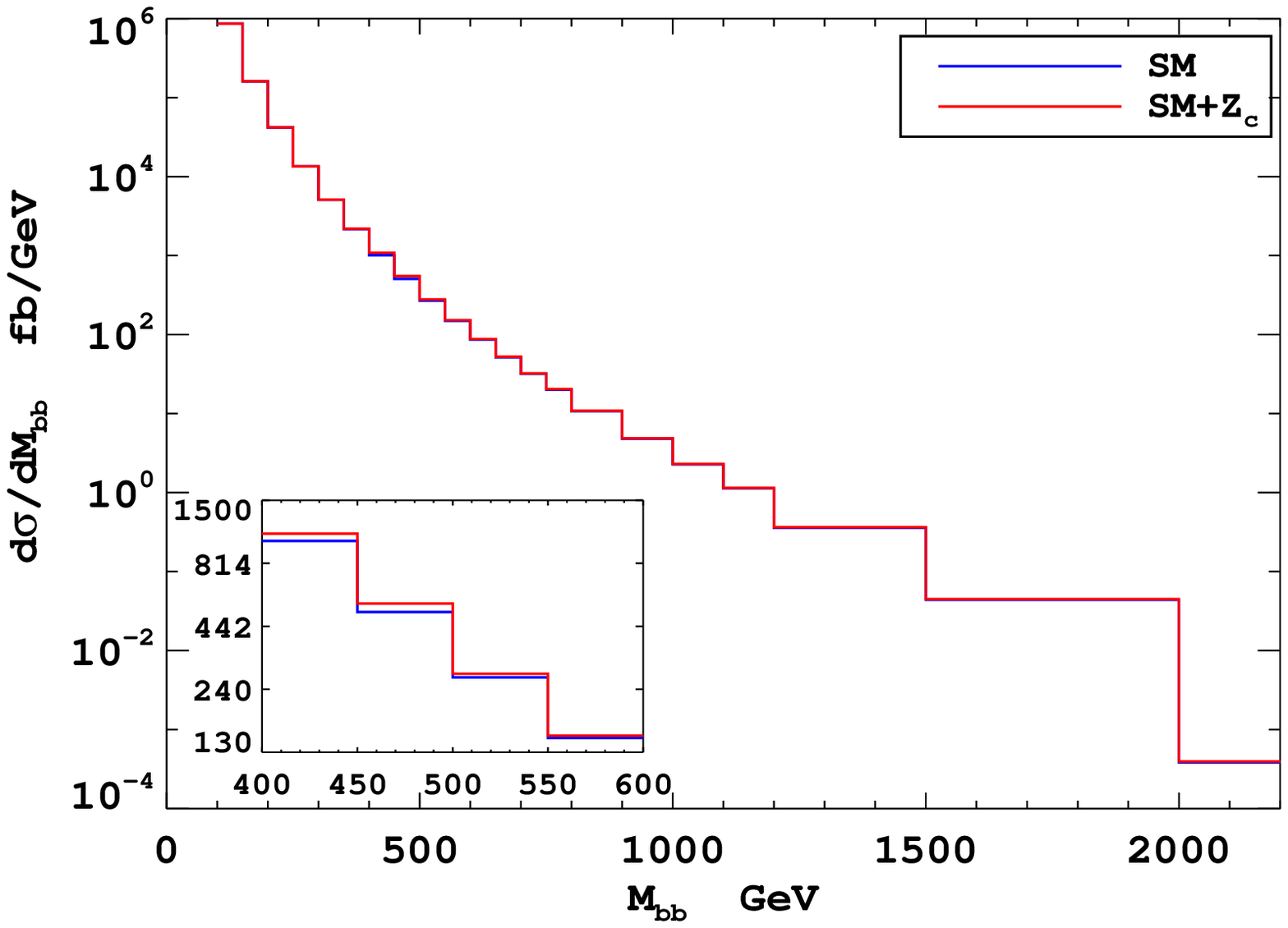}
\end{center}
\caption{\label{DCS-QQ-LHC}Differential distributions of the total
cross sections with the quark pair invariant mass. $b$ quark cut is
applied. }
\end{figure}

Figure~\ref{DCS-QQ-LHC} shows the differential distributions for the
quark pair producing cross section variate with the top and bottom
quark pair invariant mass at the LHC with $s=7~\mbox{TeV}$. For the
bottom quark, $P_T^b>20\mbox{GeV}$ and $Y<2.5$ cut is applied. Because of
the dominate proportion of the $gg\to t\bar{t}/ b\bar{b}$
channel, The bump caused by $Z_{\rm C}$ is almost completely
neglectable compared to the SM background.

\section{Conclusion and Discussion\label{sec.conclusion}}

In this paper we reexamine a color-octet pure axial-vector boson
$Z_{\rm C}$  to account for the $A_{\rm FB}$ anomaly of the $t\bar
t$ production at the Tevatron. Being a color-octet boson, $Z_{\rm
C}$ is automatically leptonhobic, free from the di-lepton final
state constraints. The pure axial-vector couplings of $Z_{\rm C}$
with quarks impact mainly on the quark angular distributions, rather
than the total cross sections. Our studies show that $Z_{\rm C}$'s
mass is about hundreds GeV and couples to light and  heavy quarks
with different strength. The best-fit parameters are $M_{\rm
C}=440~\text{GeV}$, $g_A^q=0.07$ and $g_A^Q=3$. It can account for
the measured $A_{\rm FB}$ excellently and at the same time has
little impact on the two critical observables of $d\sigma/dM_{t\bar
t}$ and dijet production. We also calculate $A_{\rm OFB}$ for top
and bottom quark pair production at the LHC, focusing on the new
contributions from $Z_{\rm C}$. Our studies show that $A_{\rm OFB}$
can be utilized  to measure the
 properties of new particle $Z_{\rm C}$.

{\em Acknowledgements:} We would like to thank Martin Schmaltz for
helpful discussions. This work is supported in part by the Natural
Sciences Foundation of China (No. 11075003).

\end{document}